\documentclass{emulateapj}

\newcommand{\etal}{{\it et al.}}
\newcommand{\kms}{km~s$^{-1}$~}

\newcommand{\be}{\begin{equation}}
\newcommand{\ee}{\end{equation}}

\begin{document}
\title{SFI++ II: A New I-band Tully-Fisher Catalog, Derivation of Peculiar Velocities and Dataset Properties}

\author{Christopher M. Springob\altaffilmark{1,2,3}, Karen L. Masters\altaffilmark{4,3}, Martha P. Haynes\altaffilmark{3,5}, Riccardo Giovanelli\altaffilmark{3,5}, and Christian Marinoni\altaffilmark{6}}

\altaffiltext{1}{Naval Research Laboratory, Remote Sensing Division, Code 7213, 4555 Overlook Avenue, SW, Washington, DC 20375; christopher.springob@nrl.navy.mil}
\altaffiltext{2}{National Research Council Postdoctoral Fellow}
\altaffiltext{3}{Center for Radiophysics and Space Research, Cornell University, Space Sciences Building, Ithaca, NY 14853; haynes@astro.cornell.edu, riccardo@astro.cornell.edu}
\altaffiltext{4}{Harvard-Smithsonian Center for Astrophysics, 60 Garden Street, Cambridge, MA 02138; kmasters@cfa.harvard.edu}
\altaffiltext{5}{National Astronomy and Ionosphere Center, Cornell University,
Ithaca, NY 14853.  The National Astronomy and Ionosphere Center is operated by
Cornell University under a cooperative agreement with the National Science
Foundation.}
\altaffiltext{6}{Centre de Physique Theorique, UMR 6207 CNRS, Universit\'e de Provence, case 907, 13288 Marseille, France; marinoni@cpt.univ-mrs.fr}

\hsize 6.5 truein
\begin{abstract}

We present the SFI++ dataset, a homogeneously
derived catalog of photometric and rotational properties and the
Tully-Fisher distances and peculiar velocities derived from them.
We make use of digital optical images, optical long-slit spectra,
and global HI line profiles to extract parameters of relevance to
disk scaling relations, incorporating several previously published
datasets as well as a new photometric sample of some 2000 objects.
According to the completeness of available redshift samples
over the sky area, we exploit both a modified percolation algorithm
and the Voronoi-Delaunay method to assign individual galaxies to groups
as well as clusters, thereby reducing scatter introduced by local
orbital motions.  We also provide corrections to the peculiar velocities for both homogeneous and inhomogeneous Malmquist bias, making use of the 2MASS Redshift Survey density field to approximate large scale structure.  We summarize the sample selection criteria,
corrections made to raw observational parameters, the grouping
techniques, and our procedure for deriving peculiar
velocities. The final SFI++ peculiar velocity catalog of
4861 field and cluster galaxies is large enough to permit the
study not just of the global statistics of large scale flows
but also of the {\it details} of the local velocity field.

\end{abstract}


\section{Introduction}

The recessional velocities of galaxies exhibit deviations from smooth Hubble flow which are induced by inhomogeneities in the matter distribution of the universe.  These deviations from smooth Hubble flow are known as peculiar velocities.  Because the peculiar velocity field is determined by the scale and amplitude of the inhomogeneities, the measurement of galaxy peculiar velocities can be used to trace the distribution of the mass, both dark and light.

The galaxy peculiar velocity field is thus a powerful tool.  It allows  us to map the large scale structures of the local universe, independent of the distribution of luminous matter.  And because the large scale structure depends on cosmological parameters, it can also be used to derive cosmological parameters, including the cosmological matter density, rate of growth of structure, and Hubble expansion rate.

The measurement of peculiar velocities is unavoidably intertwined with the measurement of distances, in that redshift-independent distances are needed in combination with redshifts to extract peculiar velocities.  Given a galaxy's observed recessional velocity $cz$ and distance $r$, as measured by a redshift-independent distance indicator, one can infer its line-of-sight peculiar velocity according to

\begin{equation}
v_{pec}=cz-H_0 r
\end{equation}

\noindent Here and in the remainder of this work, in order to avoid confusion between recessional velocities and peculiar velocities, recessional velocities are designated by $cz$, and peculiar velocities by $v$.  Both quantities will be accompanied by a subscript where appropriate.  All redshifts are assumed to be measured in the cosmic microwave background frame unless otherwise noted.

Most of the largest peculiar velocity surveys have employed secondary distance indicators---in particular the Fundamental Plane (Djorgovski \& Davis 1987, Dressler \etal ~1987; hereafter ``FP'') and Tully-Fisher relations (Tully \& Fisher 1977; hereafter, ``TF relation'').  The former expresses the luminosity of an elliptical galaxy as a power law function of its radius and velocity dispersion, while the latter expresses the luminosity of a spiral galaxy as a power law function of its rotational velocity.  The earliest surveys included peculiar velocity measurements for of order $\sim 1000$ galaxies at most.  Many of these early surveys were concatenated together into the Mark III catalog (Willick \etal ~1995, 1996, 1997).  The earliest catalogs containing more than 1000 galaxies include the Spiral Field I-band (SFI; Giovanelli \etal ~1994, 1995, Haynes \etal ~1999a, 1999b), Spiral Cluster I-band (SCI; Giovanelli \etal ~1997a, 1997b), Spiral Cluster I-band 2 (SC2; Dale \etal ~1999a, 1999b), ENEAR (da Costa \etal ~2000, Bernardi \etal ~2002), and EFAR (Colless \etal ~2000).  While these surveys have been useful for illuminating the global features of large scale motions in the universe, there are still some inconsistent results relating to fundamental aspects of the motions, such as the scale of the largest flows and the value of $\beta$ (see the review papers Dekel 1999 and Zaroubi 2001).  Larger samples are needed to resolve these discrepancies, and to investigate the {\it details} of the large scale flows.

We present here the SFI++, one of the first of a new generation of peculiar velocity surveys of $\sim 5000$ galaxies or more.  The catalog includes I-band TF parameters for 4861 field and cluster galaxies.  The only peculiar velocity catalog presently in existence to exceed the size of SFI++ is the Kinematics of the Local Universe catalog (KLUN; Theureau 1998, Theureau \etal ~1998, and references therein), which consists of 6600 galaxies with apparent B magnitudes and HI redshifts and line widths.  The B-band TF relation has significantly more scatter than the I-band relation, resulting in larger distance errors.  This is because both Galactic and internal extinction are more significant at B-band, and because the stars that supply the light are largely confined to star-forming regions---not distributed as smoothly as the stars in I-band---making accurate disk inclinations more difficult to measure.  Additionally, a significant fraction of the data in KLUN are from the literature, and are at somewhat low velocity resolution.  In contrast, the SFI++ datasets contain both new photometry derived from isophotal fitting of images and digital optical rotation curves (ORCs) and HI profiles for which methods of measuring widths have been derived specifically for the purpose of recovering the circular rotational velocity.

Observations are currently underway on the KLUN+ TF observing program, which aims to expand KLUN to include 20,000 spiral field galaxies with Nan\c cay radio telescope HI spectra, and added B and I band photometry (Theureau \etal ~2005).  Other large peculiar velocity surveys with observations in progress include the NOAO Fundamental Plane Survey (Smith \etal ~2004), which aims to provide FP measurements for 4000 early-type galaxies in 100 x-ray selected clusters within 200 Mpc/h, the 2MASS Peculiar Velocity Survey (Masters \etal ~2005), which is planned to include TF measurements for more than 5000 of the brightest inclined spirals in the 2MASS Redshift Survey (2MRS; Huchra \etal ~2005) and could be extended to include many more such objects, and the 6dF Galaxy Survey (Jones \etal ~2004), which is planned to include optical photometry and line widths, providing FP distances for 15,000 galaxies.

The TF relation requires three observational components for determining distances (and thus peculiar velocities): galaxy systemic velocities (redshifts), apparent magnitudes, and rotational velocity widths (which also require disk inclination estimates).  Our group maintains large databases of 21 cm spectral line parameters (Springob \etal ~2005, hereafter S05), optical rotation curve parameters (Catinella, Haynes, \& Giovanelli 2005) and I-band photometric parameters based on observations made over the course of the last 20+ years.  We have used subsets of these data for several past TF studies.  However, as explained in the aforementioned papers, much of this observational dataset has been newly reprocessed.  In addition, a large number of photometric observational parameters have never been published until now.  We have subsequently synthesized the entirety of these data into a new TF catalog, the SFI++, which we present here.  Because the new radio and optical spectroscopy data have already been presented by S05 and Catinella, Haynes, \& Giovanelli (2005) respectively, we have not replicated the exhaustive compilation of spectroscopic parameters here.  Instead, we include a handful of parameters most directly applicable to the computation of TF distances.  In contrast, there is no corresponding publication of our newly reprocessed photometric data, though an earlier generation of the catalog was presented by Haynes \etal ~(1999a, hereafter H99), so we provide here additional photometric parameters that are not directly used in the calculation of TF distances.

If one is to use the TF relation to compute peculiar velocities, one must first calibrate the power law with a template relation.  The SCI catalog was the first attempt by our group to derive such a template relation.  The derivation of the SCI template was presented by Giovanelli \etal ~(1997b), while the SCI dataset itself was presented by Giovanelli \etal ~(1997a, hereafter G97).  We have similarly divided our presentation of SFI++, such that the derivation of our template relation is presented in Masters \etal ~(2006; hereafter Paper I), while this work is devoted to the discussion and presentation of the catalog itself.  As explained in Paper I, a new template relation is required here because of the significantly larger data sample and the revised corrections to raw data.  But unlike G97, which just presented the cluster data, we present the data for both clusters {\it and} field galaxies in this work.  In future papers, including Springob \etal ~(2007, in prep.) and Masters \etal ~(2007, in prep.), we will use this dataset to investigate the local peculiar velocity field.

In Section 2, we review the observational selection criteria that was used for each of the individual observing campaigns that define the SFI++ sample.  In Section 3, we discuss the process by which raw observational parameters are corrected to produce physically meaningful estimates of the galaxies' luminosities and rotational velocities.  In Section 4, we present additional photometric parameters for the galaxies, separate from those directly used in the derivation of peculiar velocities.  In Section 5, we describe the cluster template we have derived in Paper I, which we use to calibrate the TF relation.  In Section 6, we describe the origin of the group assignments that we have adopted.  In Section 7, we explain the derivation of peculiar velocities for individual galaxies and galaxy groups.  The SFI++ catalog itself is presented in Section 8, and catalog properties are discussed in Section 9.  A brief summary concludes in Section 10.

\section{Selection of sample}

The complete photometric and spectroscopic data sample is composed of the previously published SFI, SCI, and SC2 datasets, and the until now unpublished Spiral Field I-band 2 (SF2) sample.  In each of these cases, prospective targets were selected from among spirals included in our private database, referred to as the Arecibo General Catalog (AGC).  We also include the TF samples of Mathewson, Ford, \& Buchhorn (1992) and Mathewson \& Ford (1996), but reprocessed to extract parameters using our methodology.

As explained in these references: the HI spectroscopy observations were made with the 305 m Arecibo telescope of the National Astronomy and Ionosphere Center\footnote{The National Astronomy and Ionosphere Center is operated by Cornell University under a management agreement with the National Science Foundation.}, the late 91 m and 42 m Green Bank telescopes of the National Radio Astronomy Observatory\footnote{The National Radio Astronomy Observatory is operated by Associated Universities, Inc., under a cooperative agreement with the National Science Foundation.}, the Nan\c cay radio telescope of the Observatory of Paris, and the Effelsberg 100 m telescope of the Max Planck Institut f\"ur Radioastronomie; the optical spectroscopic observations were made with the 2.3 m telescope at Siding Spring Observatory, the Hale 5 m telescope at Palomar Observatory\footnote{The Hale Telescope is operated by the California Institute of Technology under a cooperative agreement with Cornell University and the Jet Propulsion Laboratory.}, and the Cerro Tololo Inter-American Observatory (CTIO) 4 m telescope; and the optical photometry was done with the 1 m and 3.9 m telescopes at Siding Spring Observatory, the Kitt Peak National Observatory (KPNO) and CTIO\footnote{KPNO and CTIO are operated by Associated Universities for Research in Astronomy, under a cooperative agreement with the National Science Foundation.} 0.9 m telescopes, and the 1.3 m McGraw-Hill telescope of the Michigan-Dartmouth-MIT (MDM) Observatory\footnote{The MDM Observatory was jointly operated by the University of Michigan, Dartmouth College, and the Massachusetts Institute of Technology on Kitt Peak mountain, Arizona.}.  The details of the selection, observations, and data reduction are covered in each of the aforementioned papers.  However, below we recap some of the main features of the sample selection for each of these individual projects.

The SCI was a compilation of TF measurements for 24 clusters, compiled for the determination of a template TF relation and the determination of the motions of the clusters themselves.  The selection of the clusters used in SCI is described in G97 Section 2.3.  The clusters all have mean velocities less than 10,000 \kms when measured in the CMB frame.  They were chosen so as to span a large range in richness, and provide as much balance as possible among different parts of the sky.  As explained in G97, the SCI itself includes some data from the literature, including Pierce \& Tully (1988), Han (1992), and Han \& Mould (1992).  However, the raw observational parameters from these supplementary data were corrected using the same algorithms that were applied to the data from our own group.  All of the data were processed in the same way.

The SC2 was a TF compilation of clusters, conducted to improve the quality of the I band TF relation and to determine the redshift depth of the structures most responsible for the reflex motion of the Local Group with respect to the CMB.  5-15 TF measurements per cluster were obtained for an all-sky sample of 52 clusters with recessional velocities $5000<cz<25,000$ \kms.

The SFI was comprised of a TF sample of 2000 field galaxies limited to $cz<7500$ \kms (Local Group frame), blue magnitude $m_B<14.5$, and line width $> 100$ \kms.  The SFI also had redshift dependent upper and lower optical diameter limits, to minimize the variations in the number of objects observed per unit redshift that show up in all flux or diameter limited catalogs.  The diameter limits were $2.5'<a<5.0'$ for $cz<3000$ \kms, $1.5'<a<5.0'$ for $3000<cz<5000$ \kms, and $1.3'<a<5.0'$ for $5000<cz<7500$ \kms.

The SF2 program was intended to obtain photometry for objects either with existing HI or optical spectroscopy in our existing database at the time of the observations and to target the region $-15 < decl. < +35^\circ$ to a depth of $cz<10,000$ \kms and optical diameter $a>0.9'$.  I band images were obtained at the KPNO 0.9 m telescope.  Approximately 2300 $23' \times 23'$ fields were observed, of which roughly 1900 yielded high quality photometry.  Roughly 500 of these fields were used as part of SC2, with the remaining fields used for SF2.  The targeted fields contained at least one good TF candidate (undisturbed, inclined spiral with $cz<10000$ \kms ) for which a rotational width was already available to us from either HI line or optical long slit spectroscopy at the time the photometry was obtained.  Moreover, since additional spectroscopic observations were possible, the $23' \times 23'$ fields were centered to maximize the number of potential additional TF candidates, regardless of the status of rotational width measurements.  As a result, the number of galaxies for which photometry is available exceeds the number with rotational velocities.  Additional factors related to the allocation of telescope time, the eventual weather conditions and the practicalities of observing also had an impact on the final sample. For example, more observing time was allocated in the fall than in the spring, and the fall time had better weather. Also, while the original plan was to observe only fields with $|b|>20^\circ$, some fields closer to the Galactic plane were included to fill parts of nights when higher latitudes where not accessible.

The Mathewson, Ford, \& Buchhorn (1992) dataset was a compilation of TF measurements for 1355 Sb-Sd galaxies in the southern hemisphere, diameters $a>1.7'$, inclinations $> 40^\circ$, and Galactic latitude $|b|>11^\circ$.  Most of the objects had systemic velocities of less than 7000 \kms.  However, in the ``Great Attractor'' region, some higher redshift galaxies, drawn from the redshift survey of Dressler (1988), were included.  The Mathewson \& Ford (1996) dataset was a compilation of TF measurements for an additional 920 Sb-Sc galaxies selected from the ESO-Uppsala Survey of the ESO(B) Atlas (Lauberts 1982, hereafter ESO), with diameters $1.0'<a<1.6'$, systemic velocities between 4000 and 14,000 \kms, and the same inclination and Galactic latitude limits as the Mathewson, Ford, \& Buchhorn (1992) sample.  An additional 172 Uppsala General Catalog (Nilson 1973, hereafter UGC) galaxies were observed in the region $250<l<360^{\circ}$, $45<b<80^{\circ}$.  Both the photometric and optical spectroscopic datasets from Mathewson, Ford, \& Buchhorn (1992) and Mathewson \& Ford (1996) were made available to us and have been reprocessed using our own algorithms to achieve greater homogeneity (Haynes \etal ~1999a).

The TF sample presented here, which we refer to as SFI++, is the union of each of these datasets.  We include only those observations deemed to be of `high quality', so that, for example, HI spectral profiles assigned the quality index `G' as defined by S05 Section 4 are included, but all other HI spectra are not.  Some galaxies have photometry, but not spectroscopy (or at least not of high quality).  Others have spectroscopy but not photometry.  Such objects are excluded from consideration in the peculiar velocity catalog.  However, we do include galaxies without high quality spectroscopy in a separate photometric compilation that will be presented in Section 4.

There are also some objects with multiple photometric or spectroscopic observations.  In such cases, we choose the observations deemed to be of highest quality, and discard the others.  Preference is given to HI spectroscopy over optical spectroscopy, and to our own observations over those from the supplementary datasets.

The final sample includes 4861 galaxies for which we have good TF data, of which 807 are members of the template clusters, which we discuss in Section 5.

\section{Tully-Fisher parameters}

\subsection{Photometry}

\subsubsection{I band fluxes}

I-band optical photometric images were reduced as described by H99.  Corrected I-band apparent magnitudes and errors are then computed as described in that work.  However, while our previously published photometry used Galactic extinction corrections derived from Burstein \& Heiles (1978), we now use updated values taken from the Diffuse Infrared Background Experiment on the Cosmic Background Explorer Satellite (Schlegel, Finkbeiner, \& Davis 1998).  For estimating extinction from the target galaxy itself, we use Equation 27 from Giovanelli \etal ~(1994): $\Delta M=-\gamma$log$(a/b)$, where $\Delta M$ is the extinction in magnitudes, $a$ and $b$ are the observed semimajor and semiminor axes of the galaxy obtained from ispohote fitting, and $\gamma$ is a quantity that depends on the galaxy's inferred absolute magnitude, as described by that paper.  $\gamma$'s dependence on absolute magnitude $M_I$ is shown in Giovanelli \etal ~(1995) Figure 7c.  The exact functional form, which we adopt here, is

\begin{equation}
\gamma=0.5 ~~~{\rm for}~M_I>-19.1
\end{equation}
\begin{equation}
\gamma=1-0.417(M_I+20.3) ~~~{\rm for}~-20.3<M_I<-19.1
\end{equation}
\begin{equation}
\gamma=1.0 ~~~{\rm for}~-21.8<M_I<-20.3
\end{equation}
\begin{equation}
\gamma=1.35-0.35(M_I+22.8) ~~~{\rm for}~-22.7<M_I<-21.8
\end{equation}
\begin{equation}
\gamma=1.30 ~~~{\rm for}~M_I<-22.7
\end{equation}

We note that these relations were derived for a sample that mainly consisted of Sbc and Sc galaxies, whereas the sample presented in this paper includes a greater diversity of spiral subclasses, most notably some earlier morphological types.  One would expect that, since we use axial ratio as a tracer of inclination, and earlier morphological types tend to have thicker disks than later types, we may be underestimating the extinction in earlier types.  However, this effect should be at least partially offset by the fact that later types are likely to have more dust.  In any case, because we have derived separate morphological corrections to the TF template relation in Paper I, any underestimate of the extent to which $\gamma$ varies with morphological type should manifest itself as an increase in the TF scatter for the earlier types.  There should be no {\it systematic} effect, as the galaxies of any particular inclination are randomly distributed across the sky.

\subsubsection{Inclinations}

Once the photometric images are reduced, inclinations are then computed from the data as in G97 Section 4:  That is, from the measured semimajor and semiminor axes $a$ and $b$, we obtain the ellipticity $e=1-b/a$ as a function of the distance $r=a/2$ from the center of the galaxy.  A range of radial distances is then chosen over which the disk appears to be exponential, and we then obtain a mean value of the ellipticity $\overline{e}$.  This ellipticity is then corrected for the smearing effects of seeing, according to G97 Equation 2:

\begin{equation}
e_{corr}=1-\sqrt{(1-\overline{e})^2-\psi^2 \over 1-\psi^2}
\end{equation}

\noindent where $\psi$ is 2.5 times the size of the seeing disk for the observation, divided by the major axis at the 23.5 $mag/arcsec^2$ I-band isophote.  The inclination angle $i$ is then given by G97 Equation 3:

\begin{equation}
({\rm cos} i)^2 = {(1-e_{corr})^2-q_0^2 \over 1-q_0^2}
\end{equation}

\noindent where $q_0$ is the intrinsic axial ratio of the disk, taken to be 0.13 for galaxies of morphological type Sbc and later, and 0.20 for galaxies of earlier types.

\subsection{Spectroscopy}

\subsubsection{21 cm velocity widths}

Raw 21 cm line velocity widths were corrected for instrumental and noise effects as described in S05, Section 3.2.2.  That is, the widths are measured by taking the difference of velocities between the two midpoints of polynomials fit to either side of the line profile.  The midpoints are defined as the point at which the flux (as given by the polynomial) is $50\%$ of the {\it peak} minus {\it rms} value.  In the vast majority of cases, we used a first order polynomial.  But in rare cases, a second order polynomial was used.  Instrumental and noise corrections are applied, as is a redshift correction, all as in S05, Section 3.2.2.  6.5 \kms was then subtracted from this width as described by S05, Section 3.2.3, to produce a turbulence corrected width, which S05 refers to as $W_{c,t}$.  This value was then corrected for line-of-sight projection effects by dividing by the sine of the inclination angle {\it i}, where the inclination angle was derived as in Section 3.1.2.

In summary then, the corrected HI widths $W_{21}$ are given by

\begin{equation}
W_{21} = \left({W_{obs,21}-\Delta_s \over 1+z}-\Delta_t \right) {1 \over {\rm sin} i}
\end{equation}

\noindent where $W_{obs,21}$ is the observed width, and $\Delta_s$ and $\Delta_t$ are the instrumental and turbulence corrections respectively.  (This expression only differs from G97 Equation 5 in that we subtract the turbulence correction linearly rather than quadratically.)  As in S05, we use $\Delta_t=6.5$ \kms, and $\Delta_s=2\Delta v \lambda$, where $\Delta v$ is the spectrometer channel separation in \kms and $\lambda$ is a function of the SNR and type of smoothing as described in S05 Section 3.2.2.  See S05 Table 2 for the precise dependence of $\lambda$ on SNR and smoothing type.

Errors on $W_{21}$ are computed exactly as in G97 Equation 7, with all terms defined as in that paper.  That is, the errors are computed as the sum in quadrature of the errors on the observed width, instrumental and noise corrections, turbulence correction, and inclination correction.

\subsubsection{Optical velocity widths}

All observed optical velocity widths have been extracted by fitting a function to the folded H$\alpha$ ORCs.  We use the parametric {\it Polyex} model first described by Giovanelli \& Haynes (2002).  As described by Catinella, Haynes, \& Giovanelli (2005), this model has the functional form for the circular rotational velocity $V_{PE}$ at a distance $r$ from the galaxy's center

\begin{equation}
V_{PE}(r)=V_0 (1-e^{-r/r_{PE}})(1+\alpha r/r_{PE})
\end{equation}

\noindent $V_0$ and $r_{PE}$ are, respectively, the circular velocity amplitude and exponential scale of the inner region of the galaxy.  $\alpha$ is the slope of the outer part of the ORC.  Observed widths are measured using the value of this function at $r_{opt}$, an optical radius containing $83\%$ of the total light of the galaxy.  Corrected widths are then computed from the observed widths, as described by Catinella, Haynes, \& Giovanelli (2005), using the same cosmological and inclination corrections that are used for 21 cm line widths.

Catinella, Haynes, \& Giovanelli (2007) shows that there are slight systematic differences between widths measured from HI spectroscopy and widths measured from optical rotation curves.  These differences depend on the relative extent of the H$\alpha$ ($r_{max}$) as compared to the total optical extent of the galaxy, and on the slope of the ORC at the optical radius.  We thus correct each of the ORC widths using the relation:

\begin{equation}
W_{21}/W_{ORC}=0.899+0.188r_{max}/r_{opt} ~~~{\rm for~rising~ORCs}
\end{equation}

\begin{equation}
W_{21}/W_{ORC}=1.075-0.013r_{max}/r_{opt} ~~~{\rm for~flat~ORCs}
\end{equation}

\noindent as derived in Catinella, Haynes, \& Giovanelli (2007).  Flat ORCs are defined as those for which the gradient of the rotation curve at $r_{opt}$ is less than 0.5 \kms arcsec$^{-1}$, while rising ORCs have gradients greater than 0.5 \kms arcsec$^{-1}$.

Both optical and radio widths for which all corrections, including inclination angle corrections, have been applied are hereafter designated by `$W_{TF}$'.

\subsubsection{Recessional velocities}

For each galaxy, the recessional velocity $cz$ is taken to be the midpoint of the spectral line profile, either HI profile or ORC, as explained in S05 and Catinella, Haynes, \& Giovanelli (2005) respectively.  As in our previous TF studies, we neglect redshift errors in computing TF distances, as they are typically less than $1\%$.

\section{Additional photometric data}

We provide additional photometric parameters, of varying levels of applicability to the computation of TF distances, for an overlapping sample of 5254 galaxies.  Excluded from this overlapping photometric sample are the SC2 galaxies and much of the data we have taken from the literature, for which we are missing these additional photometric parameters.  However, we {\it include} galaxies for which we do not have high quality width measurements, which are thus not in the TF sample.

The scale length, surface magnitude, ellipticity, position angle, and observed magnitude parameters are all defined exactly as in H99, Sections 2 and 3 (with all of the same corrections applied), which itself builds on the data reduction routines developed in Giovanelli \etal ~(1994).  Note that our definition of the ellipticity is $e=1-b/a$, where $a$ and $b$ are the major and minor axes, respectively.  (In this work, unlike in H99, we use $e$ to denote ellipticity, as $\epsilon$ is used to denote errors.)

We provide these photometric parameters in Table 1.  All galaxies for which we have these additional parameters are included here.  The format of the table is as follows:

{\it Column (1).}---Entry number in the UGC, where applicable, or else in the AGC.

{\it Column (2).}---NGC or IC designation, or other name, typically from the Catalog of Galaxies and Clusters of Galaxies (Zwicky \etal ~1961-1968), ESO, or the Morphological Catalog of Galaxies (Vorontsov-Velyaminov \& Arhipova 1968).  Where used, the designation in the latter is abbreviated to eight characters.

{\it Columns (3) and (4).}---Right ascension (in {\it hh mm ss.s} format) and declination (in {\it dd mm ss} format) in J2000.0 epoch either from the NASA/IPAC Extragalactic Database (NED){\footnote {The NASA/IPAC Extragalactic Database is operated by the Jet Propulsion Laboratory, California Institute of Technology, under contract with the National Aeronautics and Space Administration.}} or measured by us on the POSS-I{\footnote{The National Geographic Society - Palomar Observatory Sky Atlas (POSS-I) was made by the California Institute of Technology with grants from the National Geographic Society.}}.  Typically, the listed positions have $\le 5$\arcsec accuracy.

{\it Column (5).}---The morphological type code following the RC3 system.  Classification comes either from the UGC or ESO catalogs, or from our own visual examination of the POSS-I prints.

{\it Column (6).}---The isophotal radius measured at an I-band surface magnitude of 23.5 mag/arcsec$^2$, $r_{23.5}$, in arcseconds.

{\it Column (7).}---The optical radius, $r_{83L}$, in arcseconds, derived from the image and corresponding to the radius encompassing $83\%$ of the light.

{\it Column (8).}---The surface magnitude at the outermost detected isophote in the I-band image, $\mu_{out}$, in mag/arcsec$^2$.

{\it Column (9).}---The disk ellipticity, $e$, corrected for seeing following Giovanelli \etal ~(1994).

{\it Column (10).}---The estimated error on the disk ellipticity, $\epsilon_e$.

{\it Column (11).}---The mean disk position angle, $PA$, in degrees, computed by averaging the position angles from a series of isophotal fits of the galaxy.

{\it Column (12).}---The estimated error on the position angle, $\epsilon_{PA}$.

{\it Column (13).}---The observed I-band magnitude, $m_{obs}$, extrapolated to 8 disk scale lengths, {\it before} the extinction and face-on corrections have been applied that produce the values used for TF.

{\it Column (14).}---The measurement error on the apparent I-band magnitude, $\epsilon_m$.

{\it Column (15).}---A code indicating whether the galaxy is included in the TF sample.  Those objects marked with a `*' are not included in the TF sample, as we do not have high quality width measurements for those galaxies.  Any object {\it not} marked with a `*' should appear in either Table 2 or Table 4.

Figure 1 shows the distribution of each of the parameters given in Table 1:  $e_0$, $r_{23.5}$, $r_{83L}$, $\mu_{out}$, $m_{obs}$, and $\epsilon_m$, for the galaxies for which we have these additional parameters.  H99 shows the histograms for exactly the same parameters for the corresponding dataset presented in that paper.  Like the H99 sample, this sample is not magnitude limited.  In fact, the distributions of $r_{23.5}$, $r_{83L}$, and $\mu_{out}$ are all remarkably similarly to the H99 distributions.  There do appear to be subtle differences in the ellipticity and magnitude error distributions however.  The typical apparent magnitude measurement error is $\sim 0.03$, as opposed to $\sim 0.04$ in H99.  Also, the ellipticities of both samples peaks at about 0.75-0.80, but the dropoff towards low inclination galaxies is more gradual in the sample presented here.  This more gradual dropoff is probably due to the inclusion of earlier morphological types, which have a larger intrinsic axial ratio, and to the fact that, unlike the SFI, SFI++ has no strict axial ratio selection criteria.

\begin{figure}
\epsscale{1.0}
\plotone{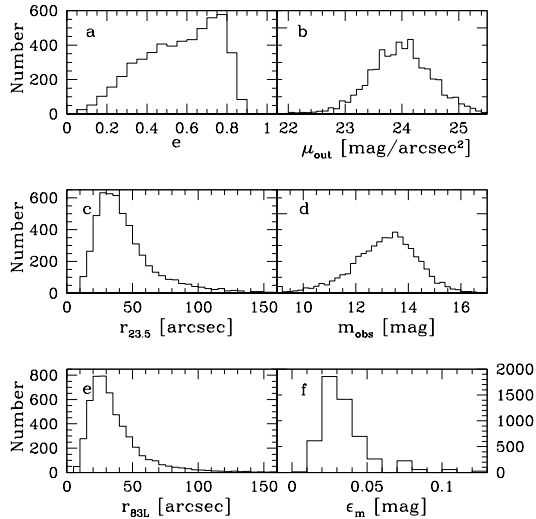}
\figcaption{Distribution of (a) ellipticity, in bins of width 0.05, (b) surface magnitude at the outermost isophote, in bins of width 0.1 magnitudes per square arcsecond, (c) isophotal radius at which the surface magnitude is 23.5 mag/arcsec$^2$, in bins of width 5 arcseconds, (d) observed I-band magnitude, in bins of width 0.2 magnitudes, (e) optical radius encompassing $83\%$ of the I-band light, in bins of width 5 arcseconds, and (f) measurement error on the apparent I-band magnitude, in bins of width 0.01 magnitudes, for the 5254 galaxies for which we have these parameters.  Note the difference in vertical scale between panels $e$ and $f$, in contrast to the identical vertical scales for the other adjacent pairs.  Figure 2 of H99 shows the distribution of the same parameters, albeit with some slightly different nomenclature, for the Sc sample that existed at that time.\label{FIG1}}
\end{figure}

\section{SFI++ cluster template}

We can express the TF relation as

\begin{equation}
L=kV_{max}^\alpha
\end{equation}

\noindent where $L$ is the luminosity of the galaxy, $V_{max}$ is the maximum rotational velocity of the galaxy, and $k$ and $\alpha$ are constants that need to be calibrated with a TF template relation.  As galaxy clusters provide a large number of objects that are located at a common distance, they are well suited for the construction of such a template.  The independent distance estimates of several galaxies in a cluster also provides a much more accurate determination of the cluster distance than is available for a single galaxy, providing us with ``hard points'' in the peculiar velocity field.  While some early TF studies depended on templates derived from a single cluster (Pierce \& Tully 1988; Mathewson, Ford, \& Buchhorn 1992), this approach offers significant drawbacks, as outlined by G97.  Giovanelli \etal ~(1997b) developed the ``basket of clusters'' method to alleviate the bulk of these problems.  In Paper I we have applied this technique to the larger SFI++ template sample to provide a new I-band TF template specifically designed for use with SFI++ galaxies.

There are 86 different clusters for which we have data in either SCI or SC2.  The cluster selection and galaxy cluster assignments are described by those papers.  We will just mention briefly that the vast majority of these are Abell clusters, though there are a few exceptions in the SCI.  As described by the aforementioned authors, cluster assignments were made by eyeball estimates based on the spatial and redshift distribution of galaxies near the cluster centers.

All of these clusters which include at least seven members in SFI++ and average systemic velocities less than 10,000 \kms in the CMB frame were considered for inclusion in the template.  While they were already used as template clusters for either SCI or SC2, the much larger data sample that we have with SFI++ compelled us to reexamine the assignments of each of the galaxies in the vicinity of these prospective template clusters.  The redshift space positions of each of the galaxies in all prospective clusters was examined by eye, just as in G97 and Dale \etal ~(1999a, 1999b).  Following the scheme of G97, we differentiate between objects believed to belong to the template cluster (the {\bf in} sample), and objects with velocities very close to the systemic velocity of the cluster but spatially removed from their center so that a firm membership assignment cannot be made (the {\bf in+} sample).  As much as possible, we have used the same thresholds for determining membership that were used for SCI and SC2.

After each prospective template cluster was examined, we discarded a few clusters because they either had too few {\bf in} objects with the new cluster assignments, or they were close enough to the galactic plane that we had concerns about differential Galactic extinction within the cluster.  We have also omitted the Virgo cluster from the template, as uncertain assignments to its various subgroups can create significant complications.  The remaining 31 clusters that match our criteria, of which 22 are from SCI and 9 are from SC2, comprise the template.  The resulting TF template relation, as derived in Paper I, expresses the corrected I band magnitude $M_I$ as a function of the corrected rotational width $W_{TF}$ as

\begin{equation}
M_I=-7.85[{\rm log}(W_{TF}) - 2.5]-20.85 + 5 {\rm log}(h).
\end{equation}

\noindent Using the terminology of Paper I, this template uses the bivariate fit to the {\bf in+} sample, with the width-dependent morphological correction applied.  That is, because of the fact that the TF relation has some dependence on morphological type, Paper I uses the template relation of Equation 14 for galaxies with morphological types of Sbc and later only, while the magnitudes of earlier type galaxies are corrected by:

$\bullet$ S0/Sa/Sab: $-0.32-0.9[{\rm log}(W_{TF})-2.5]$ mag

$\bullet$ Sb: $-0.10-0.9[{\rm log}(W_{TF})-2.5]$ mag

These corrections were derived in Paper I by fitting the offset and difference in slope between the TF relation of the later type galaxies, and the TF relations of the earlier type galaxies.  As Paper I explains, if the shallower slope of the TF relation for earlier type galaxies is intrinsic, it suggests that they have less concentrated halos than later types at a given rotational velocity.  However, the difference in slope could also be due to differing levels of incompleteness for earlier and later type spirals.

As with the SCI and SC2, the template fitting procedure also produces estimates of the peculiar velocities of the template clusters themselves.  This is also described in Paper I.

G97 explains the rationale behind the selection of the SCI/SC2 clusters, and the same principles apply here as well.  It is desirable to have `spatial balance', such that the clusters are distributed evenly across the sky, and across a range of redshifts.  This increases the probability that the `basket of clusters' is at rest with respect to the CMB.  Including clusters across a range of redshifts also allows us to sample a large dynamic range of TF parameters.

\section{Nontemplate groups}

We also exploit the group assignments of non-template galaxies so that the measured distances of each galaxy in a group can be averaged, and the TF scatter on these points can be reduced by a factor $\sim \sqrt{N}$ for a group with $N$ galaxies with TF measurements.  Group assignments for the non-template galaxies are discussed below.

Different authors have used a wide variety of group identification methods, the most common being the hierarchical (e.g., Tully 1987) and percolation (also known as ``friends-of-friends''; e.g., Huchra \& Geller 1982) methods.  The fact that both the AGC and the HI archive are inhomogeneous catalogs with no easily describable selection criteria creates significant complications for the assignment of group identifications.  However, we do have one significant advantage in this endeavor, in that we are not interested in studying the groups {\it as groups}.  We only need the groups to provide us with more accurate distance measurements.  We are thus justified in using a heterogeneous mixture of group identifications and galaxy group assignments, which we describe below.

There are three different sources from which we have drawn group assignments.  The first two, the SCI/SC2 clusters (discussed in Section 5) and Nearby Optical Galaxy (NOG) P2 groups and their extensions, have been combined into a single group catalog, which was used in the determination of the HIMF by Springob, Haynes, \& Giovanelli (2005).  The third source of group identifications, which we will discuss in Section 6.2, are Voronoi-Delaunay method (VDM) groups (Marinoni \etal ~2002).

Note: Particularly dense concentrations of galaxies are commonly referred to as clusters.  Dense clusters have markedly different properties from loose groups of galaxies, and the terms `group' and `cluster' are commonly defined in such a way that the two classifications are considered to be mutually exclusive.  However, group identification algorithms are usually designed in such a way as to identify clusters as well.  In the remainder of this work, except where stated otherwise, we define `group' broadly, so as to include clusters as well.  However, since the template clusters are all actual clusters, we continue to refer to them as `clusters'.

\subsection {NOG groups}

The basic idea involved in a percolation group-finding algorithm is presented by Huchra \& Geller (1982).  In brief: for each galaxy in the catalog in question, one searches for neighboring galaxies within a given search radius in redshift space in both the transverse and radial directions.  Any neighboring galaxies found to be within $D_L$ in the transverse direction {\it and} $V_L$ in redshift is assigned to the same group as the original galaxy.  Any such neighbors of the neighboring galaxies are also assigned to the same group, and so on.  $D_L$ and $V_L$ are referred to as `linking parameters', and their values may be fine-tuned to identify systems of a particular number density contrast.

In any flux or diameter limited survey, the density of detected objects drops off with increasing redshift.  Thus, in order for a percolation algorithm to identify groups of the same density contrast at different redshifts, the linking parameters are usually allowed to vary with redshift.

NOG is a complete, distance-limited ($cz \leq 6000$ \kms, Local Group frame) and magnitude limited ($m_B \leq 14$ mag) sample of $\sim 7000$ galaxies covering galactic latitudes $|b|> 20^\circ$ (Marinoni \etal ~1999, Giuricin \etal ~2000).  These works presented three different group catalogs with groups drawn from the NOG sample using a hierarchical grouping algorithm and two different variants of a percolation algorithm: one with the linking parameters kept constant, and the other with the linking parameters scaled with redshift.

We have cross-referenced the AGC with NOG galaxies with group identifications in the percolation group catalog that has redshift dependent linking parameters, referred to by Giuricin \etal ~as the `P2' group catalog.  We have combined the P2 group identifications with the SCI/SC2 group assignments described in the previous section, to generate a single catalog of group assignments.

Because in most of the sky, the AGC goes significantly deeper than NOG's 14 mag limit, we have also developed a modified percolation algorithm that extends NOG groups to include nearby AGC galaxies that are too dim to be included in the original NOG catalog.  The AGC is nearly complete down to $m_B=15.7$ north of $decl.=-2^\circ$, and nearly complete to $m_B=15.0$ south of $decl.=-2^\circ$, but with a significant number of galaxies dimmer than these completeness levels as well.  So we have adjusted the linking parameters to represent the same level of density contrast as NOG, but with these deeper magnitude limits---the relationships between the density contrast and the linking parameters are explained in Section 2 of Huchra \& Geller (1982).  We then made slight adjustments to the parameters to include the AGC galaxies that are even dimmer than the AGC magnitude limits.  The new linking parameters were then used to add AGC galaxies to nearby NOG groups, but {\it not} to generate new groups.  This procedure is explained in more detail by Springob (2006).

The combined group catalog of SCI/SC2 clusters and NOG P2 groups with extensions is used for all SFI++ group assignments south of $decl.=-2.0^\circ$.

\subsection {Voronoi-Delaunay groups}

The catalog described above is useful for numerous applications, however it also has some significant drawbacks.  First, the NOG groups extend to a systemic velocity of only 6000 \kms.  The AGC includes a significant number of galaxies beyond this velocity.  And second, the percolation method for generating group assignments has several drawbacks, as pointed out by Marinoni \etal ~(2002): The technique is insensitive to local variations in the density of points, the fine tuning of the linking parameters may lead to systematic differences in the properties of groups identified at different redshifts in the same dataset, and the technique may also identify ``groups'' that are not physically realistic---physically distinct concentrations of galaxies connected by long chains of galaxies.

It is precisely such drawbacks inherent in percolation grouping that motivated Marinoni \etal ~(2002) to propose a more physically motivated group-finding algorithm, the Voronoi-Delaunay method (VDM).  The Voronoi partition, by non-parametrically smoothing data, represents an efficient way to measure packing and identify as potential group centers the density peaks in the galaxy distribution, while the Delaunay mesh, by reconstructing the neighborhood relationship between galaxies, represents a natural way to assign group members.

The procedure works as follows: Galaxy group centers are identified by peaks in the galaxy density field.  All galaxies located within a cylindrical volume in redshift space centered on that concentration are assigned to the group.  According to this strategy, there is no need to introduce an arbitrarily chosen {\it global} density threshold to judge when a given system is formed.  Instead, the dimensions of the cylinder are {\it locally} scaled on the basis of physical considerations, i.e., according to the richness-velocity dispersion correlation (e.g., Bahcall 1981).  The details of the relationship between these dimensions and the group richness are derived from semianalytic galaxy formation algorithms applied to N-body simulations.  The method is described in detail by Marinoni \etal ~(2002), and was used to identify groups in the DEEP2 Galaxy Redshift Survey (Gerke \etal ~2005).

North of $decl.=-2.0^\circ$, the AGC is nearly complete in optical diameter to $1.0'$ outside the zone of avoidance (ZOA).  We have artificially filled in the ZOA of the AGC to eliminate edge effects, and then used C.M.'s VDM code to generate VDM group assignments for AGC galaxies with $a>1.0'$ and $decl.>-2.0^\circ$.

For the simulation of large scale structure in the ZOA, we first discarded all real galaxies with $|b|<15^\circ$, then filled in that region by creating duplicates of all galaxies in the region $-2.0<decl.<40.0^\circ$, but shifted by $+5^h$ in R.A, and duplicates of all galaxies north of $decl.=40.0^\circ$, but shifted by $12^h$ in R.A.  Any duplicate galaxies lying {\it within} $|b|<15^{\circ}$ after having been shifted in R.A. is retained as a `synthetic galaxy' in the ZOA.  All duplicates outside of that region are discarded.

The VDM algorithm operates on real and synthetic galaxies alike, and so several groups are composed exclusively of synthetic galaxies, or of a combination of real and synthetic galaxies.  This is necessary for the initial group identification.  But the final group catalog is purged of all synthetic galaxies, and only groups with at least two real galaxies have been retained.  

The catalog contains 5423 real galaxies in 1071 groups of 2 or more real galaxies.  However, only 355 of those groups contain 5 or more members.  Like the AGC, the mean systemic velocities of the vast majority of the groups is less than 10,000 \kms.  However, there are some groups with greater distances, including one binary pair with a mean systemic velocity of nearly 22,000 \kms.

\subsection{Combined catalog and group statistics}

We emphasize that all template galaxies retain their template group assignments regardless of any group assignments in the VDM catalog or previous assignments in the SCI/SC2/NOG catalog.  For all other SFI++ galaxies, we use SCI/SC2/NOG group assignments for objects south of $decl.=-2^{\circ}$ or systemic velocities of $cz<1000$ \kms and VDM group assignments for all objects outside of those limits.

For this hybrid group catalog, there are 1360 SFI++ galaxies in 736 non-template groups.  However, the majority of these groups include only one SFI++ galaxy.  In these cases, we still use the group redshift as opposed to the galaxy redshift in computing the peculiar velocity, in order to remove the effects of any small scale motions of the galaxies with respect to the group center of mass.  Among the 288 groups that include more than one SFI++ galaxy, only 22 contain more than 5 SFI++ galaxies.

In order to provide the interested reader with an idea of the sizes of the groups involved, the data tables presented in this work also include the total number of AGC galaxies in each group.  However, we caution that these numbers should not be used for quantitative applications.  They are only meant as a guide for the reader to distinguish between large groups and small groups.  The numbers of galaxies in the VDM groups does have some significance because of the way the VDM grouping was done.  But because of the inhomogenous nature of the AGC, the galaxy counts in the other groups is less meaningful.

Also, as a rough guide to the uncertainty in the groups' systemic velocities, we provide a group redshift error statistic, derived under the assumption that each galaxy redshift serves as an independent measurement of the group redshift, neglecting the error introduced by the fact that there is some chance that a galaxy's group assignment may be erroneous.  The error, $\epsilon_{cz}$, on the CMB frame velocity, $cz_{group}$, of a group with $N_{AGC}$ AGC galaxies is then given by the sum in quadrature of the errors on CMB frame velocities of the individual galaxies:

\begin{equation}
\epsilon_{cz}={\sqrt{N_{AGC}} \over \sum_{i=1,N_{AGC}} 1 / \epsilon_i}
\end{equation}

\noindent where $\epsilon_i$ is the measurement error on the CMB frame velocity for each individual galaxy in the group.  This measurement error is typically just a few \kms for radio redshifts, but can be $\sim 30$ \kms for optical redshifts.  The AGC includes some optical redshifts for which this error is unknown.  In such cases, we conservatively assume $\epsilon_i=40$ \kms.

This grouping scheme represents a unique feature of SFI++, as past peculiar velocity catalogs have largely been restricted to either `cluster' or `field' samples, with no provision made for groups that have only a handful of members with peculiar velocity data.  Exploiting such grouping data is possible because of the large number of objects in SFI++, which exceeds that of almost every past peculiar velocity catalog.  Willick \etal ~(1996) made the only previous attempt to incorporate galaxy grouping information into the construction of peculiar velocity catalogs.

\section{Derivation of peculiar velocities}

\subsection{Peculiar velocities of field galaxies}

Peculiar velocity estimates for galaxies not assigned to any group are made according to

\begin{equation}
v_{gal}=cz_{gal}(1-10^{0.2dm})
\end{equation}

\noindent which means that the distance to the galaxy (in velocity units) can then be expressed as

\begin{equation}
r_{gal}=cz_{gal}-v_{gal}
\end{equation}

\noindent where $cz_{gal}$ is the systemic velocity of the galaxy, and $dm$ is the difference between the corrected absolute magnitude of the galaxy and the predicted absolute magnitude that one would expect from the template, given that galaxy's corrected width measurement.  In computing $dm$ for individual galaxies, we also make use of the same width-dependent morphological correction for earlier types that is described in Section 5.  This treatment of the magnitude as a function of the width (rather than the other way around) is known as the ``forward TF relation''.

Peculiar velocity errors are taken as the sum in quadrature of the absolute magnitude error, velocity width error, and an intrinsic TF scatter term.  The intrinsic scatter, in magnitude units, is given by

\begin{equation}
\epsilon_{int}=0.35-0.37 [{\rm log}(W_{TF})-2.5]
\end{equation}

\noindent as derived in Paper I.

\subsection{Peculiar velocities of groups}

For each group with more than one galaxy in the sample, we determine group peculiar velocities by averaging the peculiar velocities of each group member, weighted by the error on each individual peculiar velocity.  That is, a group with $N_{SFI++}$ galaxies with peculiar velocities $v_i$ and velocity errors $\epsilon_i$ has velocity

\begin{equation}
v_{group}={\sum_{i=1,N_{SFI++}} v_i / \epsilon_i \over \sum_{i=1,N_{SFI++}} 1 / \epsilon_i}
\end{equation}

\noindent where $v_i$ is computed slightly differently from the method explained in Section 7.1.  Instead of using the galaxy redshift as in Equation 16, we use the group redshift $cz_{group}$, so that

\begin{equation}
v_i=cz_{group}(1-10^{0.2dm})
\end{equation}

\noindent where $cz_{group}$ is the mean redshift of all galaxies in the group, including those AGC galaxies for which we have redshift data, but not sufficient photometric or spectroscopic data to include in SFI++.

Using the same mathematical reasoning as in Equation 15, the group peculiar velocity error is then given by

\begin{equation}
\epsilon_{group}={\sqrt{N_{SFI++}} \over \sum_{i=1,N_{SFI++}} 1 / \epsilon_i}
\end{equation}

\subsection{Malmquist bias}

`Malmquist bias' is the term generally used to refer to biases originating from the spatial distribution of objects (Malmquist 1924).  It arises from the coupling between the random distance errors and the density variation along the line of sight.  Because of these density variations, the probability distribution for the distance cannot simply be modeled by a gaussian along the line of sight, centered on the measured distance.

There are two types of Malmquist bias that one must be concerned with.  First, for a given set of selection criteria, the probability of one's sample including a galaxy with a given apparent magnitude will vary with the distance to that galaxy, as a result of several factors: 1) the luminosity function is not perfectly flat, 2) within a given solid angle, there are more galaxies at larger distances than smaller distances, and 3) the selection function for the galaxies may vary with distance and/or redshift.  While this bias is homogeneous across the sky, however, the second form of Malmquist bias is inhomogeneous.  It arises from the variations in large scale structure along the line of sight.  Failure to account for {\it this} type of bias can lead to spurious infall signatures onto high density regions.  Examples of Malmquist bias correction schemes for peculiar velocity samples include Freudling \etal ~1995 and Park \& Park 2006, which correct for Malmquist bias in the SFI catalog.

While it is fairly straightforward to correct for both the inhomogeneous Malmquist bias resulting from large scale structure (provided one has access to a reconstruction of the local density field at hand) and the homogeneous Malmquist bias resulting from the volume effect and the luminosity function, one still needs to account for the selection criteria of the sample.  This is a serious problem for our dataset, in that our selection criteria are very inhomogeneous.  The best we can hope to do is to construct ad hoc selection criteria that will mimic the observational properties of the catalog, without regard to the prior selection criteria that were used to generate the sample.  

We have thus adopted a procedure in which we assume that, whatever the underlying luminosity function and sky distribution of sources may be, our selection function forces the output catalog to have the particular magnitude distribution that is observed at each redshift.  That is, whatever the underlying distribution of galaxies that could be used as TF targets, our selection criteria chooses targets according to a probability function that must result in the final catalog having precisely the same bivariate distribution of redshift and magnitude that our catalog has.  We can then compute the probability, $p(r)$, that a given galaxy is at a distance $r$ by convolving the a priori TF-measured distance and associated errorbar with the number of galaxies per magnitude bin at each distance (this takes care of the homogeneous Malmquist bias) and the large scale density field (this takes care of of the inhomogeneous Malmquist bias).

We emphasize that this approach has been adopted by necessity, owing to our inhomogeneous selection criteria.  If the reader wishes to make use of a subsample of our data for which a homogeneous selection function can be applied, we strongly recommend adopting a different Malmquist bias correction approach that makes use of those selection criteria.  Additionally, if one were to extract a subset that imposed additional criteria, such as excluding galaxies outside of a given distance or redshift range, one would need to impose an additional bias correction.  So we urge extreme caution in using these Malmquist bias-corrected distances.  This is in fact why we are providing both corrected and uncorrected distances---so that the reader will have the ability to use his or her own correction procedure if he or she wishes to.  Having said that, we describe the details of this bias correction procedure below.

\subsubsection{Bias correction procedure}

We break up SFI++ into three declination bins, $decl.<-17.5^\circ$, $-17.5<decl.<-2.5^\circ$, and $decl.>-2.5^\circ$, I band apparent magnitude bins of width 0.5 mag, and redshift bins of width 1000 \kms .  (The declination bins are used because of the variation in completeness with declination.  {\it Within} each declination bin, the variation in selection criteria from one point on the sky to the next is negligible.)  We then make the assumption that the apparent magnitude distribution within each bin of declination and {\it systemic velocity} is most likely to be the same as the apparent magnitude distribution within a corresponding bin of declination and {\it distance}.  (This is likely to be approximately true as, averaged across the entire sky, a particular distance bin should be expected to have a roughly equal number of galaxies with positive and negative peculiar velocities.)

For each galaxy, we now convolve the probability distribution $p_{TF}(r)$ of its distance from the TF measurement with the probability $p_{mag}(r)$ of finding a galaxy with its apparent magnitude at each possible distance.  Further, in order to account for the inhomogeneous Malmquist bias resulting from large scale structure, we convolve these probability distributions with the density field reconstructed from the 2MRS (Erdogdu \etal ~2006), provided to us by Pirin Erdogdu.  The probability of the corrected distance to the galaxy being $r_i$ is then

\begin{equation}
p(r_i)=k_1 p_{TF}(r_i)p_{mag}(r_i)p_{lss}(r_i)
\end{equation}

\noindent where $p_{lss}(r_i)$ is the density distribution along the line of sight for the galaxy, as given by the 2MRS density field, and $k_1$ is a normalization constant, such that $\Sigma_i p(r_i)=1$.  For a galaxy with TF-measured distance $r_{gal}$ and peculiar velocity error $\epsilon_v$, the summation is done over 41 different distances ranging from $r_{gal}-2\epsilon_v$ to $r_{gal}+2\epsilon_v$, separated by $0.1\epsilon_v$.  $p(r_i)$ is evaluated at each of those distances, and the Malmquist bias corrected distance is then given by

\begin{equation}
r_{gal-malm}=\Sigma_{i=1,41} p(r_i)r_i
\end{equation}

\noindent with Malmquist bias-corrected peculiar velocity

\begin{equation}
v_{gal-malm}=cz_{gal}-r_{gal-malm}
\end{equation}

\noindent and distance / peculiar velocity error

\begin{equation}
\epsilon_{v-malm}=k_2 \sqrt{\Sigma_{i=1,41} p(r_i)(r_i-r_{gal-malm})^2}
\end{equation}

\noindent where $k_2=1.35$ is a coefficient that we include in order to correct for the fact that we are evaluating $p(r)$ only within the range of $r_{gal}-2\epsilon_v$ to $r_{gal}+2\epsilon_v$.

This is the correction that we apply {\it for field galaxies}.  As explained in Paper I, the Malmquist bias for the template clusters is likely to be negligible, so we ignore it.  However, it should be applied for the field galaxies and nontemplate groups.  For the groups, we also provide Malmquist bias corrected peculiar velocities and errors.  We use the same relations that are shown in Equations 19 and 21, but substitute in the Malmquist bias corrected velocities and errors for the individual galaxies.  So, a group with $N_{SFI++}$ galaxies with Malmquist bias corrected peculiar velocities $v_{malm,i}$ and errors $\epsilon_{malm,i}$ would have a group velocity

\begin{equation}
v_{group-malm}={\sum_{i=1,N_{SFI++}} v_{malm,i} / \epsilon_{malm,i} \over \sum_{i=1,N_{SFI++}} 1 / \epsilon_{malm,i}}
\end{equation}

\noindent where $v_{malm,i}$ is computed according to Equation 24, but again (as in the case of the uncorrected group velocities) using the group redshifts rather than individual galaxy redshifts.

Our Malmquist bias corrected group velocity error also follows Equation 21 closely:

\begin{equation}
\epsilon_{group-malm}={\sqrt{N_{SFI++}} \over \sum_{i=1,N_{SFI++}} 1 / \epsilon_{malm,i}}
\end{equation}

One final note about this procedure must be made.  The 2MRS density field only extends out to 20,000 \kms .  Additionally, our magnitude and redshift binning does not extend beyond 20,000 \kms , as the number of galaxies becomes too sparse at that point.  We thus assume a flat magnitude distribution and a flat density distribution beyond 20,000 \kms .  Therefore, {\it our Malmquist bias corrections for galaxies with distances close to or greater than 20,000 \kms tend to represent little improvement over the uncorrected distances, and should be taken with a large grain of salt.}  There are, in any case, very few SFI++ galaxies with distances in that regime.

\section{TF Data Compilation}

Here, we provide the corrected photometric and spectroscopic parameters directly used to compute the peculiar velocities, as well as the peculiar velocities themselves.  We are also making these data available online in US National Virtual Observatory (NVO)-compliant tables hosted by the Cornell Theory Center\footnote{see {\it http://arecibo.tc.cornell.edu/hiarchive/sfiplusplus.php}}.  In Table 2, we present the observational parameters and peculiar velocities for all galaxies that are not in any of the template clusters.  This includes galaxies that we identify as belonging to nontemplate groups, as described by Section 6.  In Table 3, we present observational parameters and peculiar velocities for all nontemplate groups.  And in Table 4, we present observational parameters for all galaxies in the template clusters.  Parameters for the template clusters themselves can be found in Paper I.

The format of Table 2 (all galaxies not in the template sample) is the same as Table 1 for the first five columns.  We then add the following parameters:

{\it Column (6).}---Logarithm of the corrected rotational velocity width in units of \kms, log($W_{TF}$).

{\it Column (7).}---The estimated error on log($W_{TF}$), $\epsilon_w$.

{\it Column (8).}---The observed I-band magnitude, $m_{obs}$, extrapolated to $8r_d$, {\it before} the extinction and face-on corrections have been applied that produce the values used for TF.  (This is the same parameter as Table 1, Column 13.)

{\it Column (9).}---The apparent magnitude $m_I$, corrected for extinction, and corrected to face-on magnitude.

{\it Column (10).}---The corrected absolute magnitude $M_I$, computed assuming that the galaxy is at the distance given by its redshift, with $H_0=100$ \kms.

{\it Column (11).}---The estimated error $\epsilon_M$ on the absolute magnitude.  This is not to be confused with the {\it apparent} magnitude error given in column (14) of Table 1, which does not include the error contributions from the uncertainty in the inclination, redshift, etc.

{\it Column (12).}---The extinction coefficient $\gamma$, in magnitudes, computed according to Equations 2-6.

{\it Column (13).}---The inclination $i$ of the plane of the disk to the line of sight, in degrees.

{\it Column (14).}---The velocity of the galaxy in the CMB frame, $cz_{gal}$, in \kms, taken to be the midpoint of the spectral line profile, regardless of whether optical or radio spectroscopy was used.

{\it Column (15).}---The peculiar velocity $v_{gal}$ in \kms, uncorrected for Malmquist bias, as given by Equation 16 for field galaxies.  For galaxies in groups, we provide the peculiar velocity estimate before averaging by groups, as given by Equation 20.

{\it Column (16).}---The estimated error on the (uncorrected for Malmquist bias) peculiar velocity, $\epsilon_v$, in \kms.

{\it Column (17).}---The (uncorrected for Malmquist bias) distance to the galaxy, $r_{gal}$, as given by $cz_{gal}-v_{gal}$ as in Equation 17, in \kms, for field galaxies.  For galaxies in groups, we provide the distance estimate before averaging by groups, $cz_{group}-v_{gal}$.

{\it Column (18).}---The Malmquist bias corrected peculiar velocity $v_{gal-malm}$ in \kms, as given by Equation 24 for field galaxies.  For galaxies in groups, we provide the peculiar velocity estimate before averaging by groups.

{\it Column (19).}---The estimated error on the Malmquist bias corrected peculiar velocity, $\epsilon_{v-malm}$, in \kms.

{\it Column (20).}---The Malmquist bias corrected distance to the galaxy, $r_{gal-malm}=cz_{gal}-v_{gal-malm}$, in \kms, for field galaxies.  For galaxies in groups, we provide the distance estimate before averaging by groups, $cz_{group}-v_{gal-malm}$.

{\it Column (21).}---Group number that the galaxy is assigned to.  For SCI/SC2 clusters, we use the same numbering scheme used for SCI/SC2.  For NOG groups, we use the NOG P2 numbering scheme, but with 30,000 added to each group number, so that any group number between 30,000 and 40,000 is a NOG group.  For VDM groups, the numbering scheme begins with group number 40,001, so that any group number greater than 40,000 is a VDM group.  If the galaxy does not have a group assignment, the number is given as `0'.

{\it Column (22).}---A code indicating whether the spectroscopic data are radio or optical.  `H' for HI spectroscopy and `O' for optical spectroscopy.

The format of Table 3 (non-template groups) is as follows:

{\it Column (1).}---Group number, using the same numbering scheme as Table 2, column (21).

{\it Columns (2) and (3).}---Right ascension (in {\it hh mm ss.s} format) and declination (in {\it dd mm ss} format) in J2000.0 epoch of the group center.

{\it Column (4).}---The CMB frame velocity, $cz_{group}$, in \kms, taken to be the average of the CMB velocities of the constituent galaxies, regardless of their inclusion in SFI++.

{\it Column (5).}---The `error' on the CMB frame velocity, $\epsilon_{cz}$, in \kms, computed according to Equation 15.

{\it Column (6).}---The number of galaxies in the group, $N_{SFI++}$, for which we have data in SFI++.

{\it Column (7).}---The number of AGC galaxies in the group, $N_{AGC}$.

{\it Column (8).}---The group peculiar velocity $v_{group}$, in \kms, uncorrected for Malmquist bias, computed according to Equation 19.

{\it Column (9).}---The estimated error on the (uncorrected for Malmquist bias) group peculiar velocity, $\epsilon_{group}$, in \kms, computed according to Equation 21.

{\it Column (10).}---The (uncorrected for Malmquist bias) distance to the group, $r_{group}$, as given by $cz_{group}-v_{group}$, in \kms.

{\it Column (11).}---The Malmquist bias corrected group peculiar velocity $v_{group-malm}$, in \kms, uncorrected for Malmquist bias, computed according to Equation 26.

{\it Column (12).}---The estimated error on the Malmquist bias corrected group peculiar velocity, $\epsilon_{group-malm}$, in \kms, computed according to Equation 27.

{\it Column (13).}---The Malmquist bias corrected distance to the group, $r_{group-malm}=cz_{group}-v_{group-malm}$, in \kms.

The format of Table 4 (galaxies in the template sample) follows that of Table 2, except that we omit columns 15-22 and add columns (23), the angular separation, $\theta$, between the galaxy and the cluster center, in units of arcminutes, and (24), the template cluster name, using the same naming scheme as in Paper I.  We have also added column (25), which contains both the spectroscopic data code from Table 2, column (22), and a second code which characterizes membership status in the cluster.  As in G97, code `c' signifies a {\it bona fide} cluster member (the {\bf in} sample), while code `g' indicates that a firm membership assignment cannot be made (the {\bf in+} sample).  One other difference between Tables 2 and 4 is that, following G97, we use the cluster redshift to compute the absolute magnitude of {\bf in} galaxies in column (10).  For {\bf in+} galaxies, we use the galaxy redshift, as in Table 2.

\section{Dataset Characteristics}

The sky distribution of all SFI++ galaxies is shown in Figure 2.  As with the S05 HI archive, very few objects can be found close to the Galactic plane.  SFI++ does, however, include a far greater share of galaxies in the southern hemisphere than the HI archive; however there is still a deficiency of galaxies in the range $-17.5<decl.<-2.5^\circ$, owing to the fact that all of our observing programs targeted known galaxies in published catalogs, and there is no catalog of comparable depth to the UGC or ESO in this declination range.  We also note that, due to the large amount of HI data we have in the Arecibo declination range, there is better coverage there ($-2<decl.<+38^\circ$) than in any other part of the sky.  In Figure 3, we show the sky distribution of the template clusters.  The clusters are broadly spread across the sky, though there is a greater concentration in both the Pisces-Perseus Supercluster and the supergalactic plane.

\begin{figure}
\epsscale{1.0}
\plotone{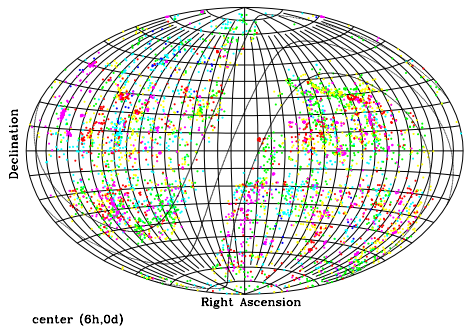}
\figcaption{Aitoff equal area projection of the sky distribution of SFI++ galaxies.  The plot is centered at R.A. = $6^h$.  The thick lines 
trace the galactic latitudes $b=-20^\circ$, $b=0^\circ$, 
and $b=+20^\circ$.  The figure is color-coded by CMB frame redshift, such that $cz_{gal}<1000$ \kms galaxies are blue, $1000<cz_{gal}<3000$ \kms galaxies are cyan, $3000<cz_{gal}<5000$ \kms galaxies are green, $5000<cz_{gal}<7000$ \kms galaxies are yellow, $7000<cz_{gal}<9000$ \kms galaxies are red, and $cz_{gal}>9000$ \kms galaxies are magenta.\label{FIG2}}
\end{figure}

\begin{figure}
\epsscale{1.0}
\plotone{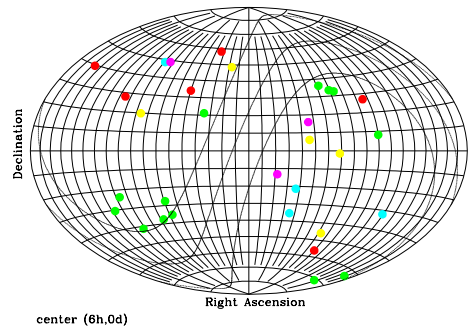}
\figcaption{Same as Figure 2, but for SFI++ template clusters.  We use the same coloring scheme as in Figure 2, color-coded by the mean cluster redshift.\label{FIG3}}
\end{figure}

Figure 4 shows the distributions of apparent and absolute magnitudes, widths, radial velocities, and TF distances, both corrected and uncorrected for Malmquist bias.  Because of the redshift-dependent diameter limit of the SFI, specifically designed to equalize the number of objects across a large range of distances, the radial velocity distributions of Figure 4d is somewhat ``flatter'' than the corresponding radial velocity distribution of the HI archive shown in S05 Figure 4.  As in that figure, the redshift distribution peaks at $\sim 5000$ \kms, which corresponds to the mean recessional velocity of the densest concentration of the Pisces-Perseus Supercluster.  Because the TF distances for individual galaxies have errors of $\sim 15\%$, that redshift peak is even more smoothed out in the TF distance histogram, though moreso for the histogram of uncorrected distances (Fig. 4e) than the histogram of Malmquist bias corrected distances (Fig. 4f).

\begin{figure}
\epsscale{1.0}
\plotone{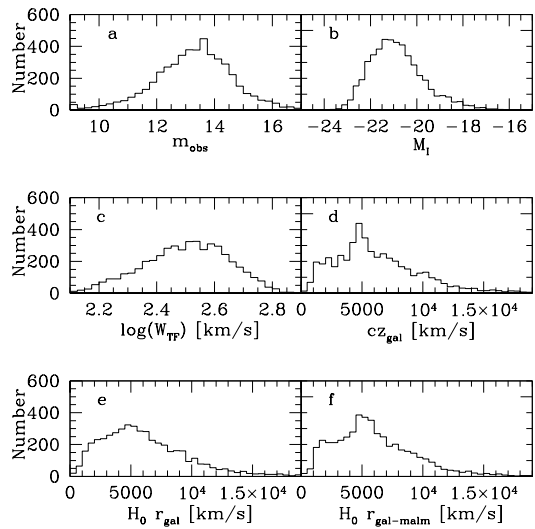}
\figcaption{Distribution of (a) Observed I band magnitude, in bins of width 0.25 mag, (b) I band absolute magnitude, in bins of width 0.25 mag, (c) logarithm of velocity widths, in bins of width 0.025 dex, (d) recessional velocity in CMB frame, in bins of width 500 \kms , (e) TF distance (uncorrected for Malmquist bias), in bins of width 500 \kms, and (f) Malmquist bias corrected TF distance, in bins of width 500\kms, for all galaxies in the SFI++.\label{FIG4}}
\end{figure}

The distributions of the peculiar velocities of individual galaxies, nontemplate groups, and template clusters are shown in Figure 5.  (In the case of individual galaxies and nontemplate groups, we provide the distributions of peculiar velocities in both the Malmquist bias corrected and uncorrected cases.)  We emphasize that very few galaxies have {\it real} peculiar velocities of magnitude greater than 1000 \kms.  In fact, few galaxies outside of clusters have real peculiar velocities of magnitude greater than 500 \kms.  Thus, the broad distribution of peculiar velocities for individual galaxies and small groups is a consequence of the scatter in the TF relation---some galaxies are intrinsically offset from the TF relation, and some simply have unusually large magnitude or width errors.

\begin{figure}
\epsscale{1.0}
\plotone{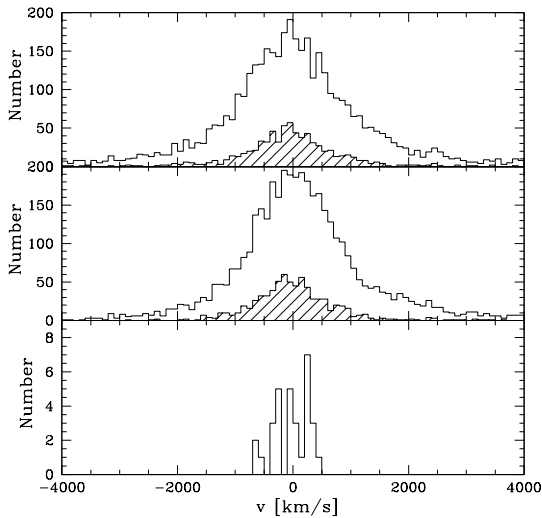}
\figcaption{{\it Top} Peculiar velocity distribution of all nontemplate groups (hatched) and all field galaxies (unhatched) {\it before} the Malmquist bias correction; {\it middle} peculiar velocity distribution of all nontemplate groups (hatched) and all field galaxies (unhatched) {\it after} the Malmquist bias correction; {\it bottom} peculiar velocity distribution of the template clusters.  In all three plots, the bins width is 100 \kms.\label{FIG5}}
\end{figure}

The peculiar velocities of template clusters fall in a much narrower range, both because of the reduced distance errors associated with these clusters, and the fact that larger groups and clusters tend to have smaller peculiar velocities than do field galaxies or small groups.  The dispersion of template cluster peculiar velocities is discussed in Paper I.

The TF relation itself is plotted in Figure 6, superimposed against Paper I's template relation.  All nontemplate galaxies are included in this figure, with the magnitudes for earlier type galaxies corrected by the width-dependent morphological corrections of Section 5.  Because this plot does not account for the Malmquist bias correction, selection effects should be apparent, and we expect there to be some deviation from the template relation.  However, there are some extreme outliers---many of which have unusually large width errors.  Most, but not all, of these galaxies with large width errors have small inclination angles, so the corrections to edge-on widths are large.  Some simply have larger-than-average measurement errors---these are far more likely to be optical widths than radio widths.  There do remain some outliers, however, that do not have unusually large magnitude or width errors.  It is possible that in such cases, we have underestimated the uncertainties.  One complicating factor is that, since the TF relation depends on morphological type, ambiguity in morphological classification will inevitably introduce its own sources of error, though this error is extremely difficult to quantify.  This problem is more significant for this dataset than in most of our group's previous TF studies, because SFI++ includes more galaxies of smaller optical extent, which are more difficult to classify.  In any case, the extreme outliers from the TF relation would likely be excluded for most applications of the catalog, but we leave it up to the reader to decide on an appropriate cutoff.

\begin{figure}
\epsscale{1.0}
\plotone{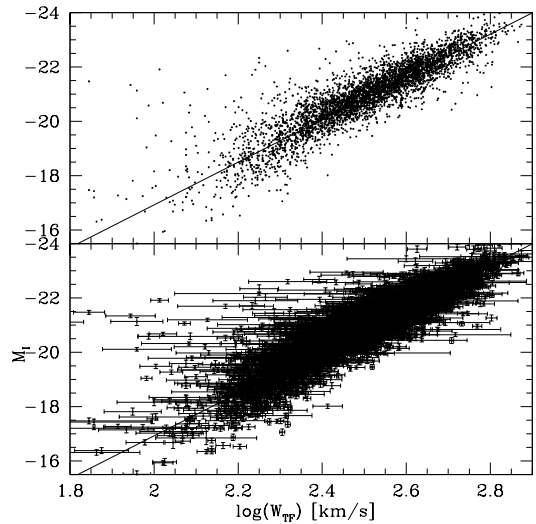}
\figcaption{{\it Top} TF relation for all nontemplate galaxies.  {\it Bottom} Same points plotted, but with magnitude and width errorbars included.  The solid line represents the template relation from Paper I, which we provide in Equation 14.  The width-dependent morphological corrections of Section 5 have been applied to the magnitudes plotted here, {\it unlike} the magnitudes in Table 2 and 4. \label{FIG6}}
\end{figure}

\section{Summary}

We have constructed a TF catalog, the SFI++, including 4861 field and cluster galaxies, primarily based on observations made by our group and collaborators over the course of the last $\sim 25$ years, but also including some data from the literature that has been processed in an identical manner.  While most of the spectroscopic data have already been published, all of the data have now been reprocessed using new correction schemes.  And a significant fraction of the photometric data are provided here for the first time.  We have reexamined the choice of clusters used in the construction of a TF template, as well as the group assignments of all galaxies in the vicinities of those clusters.  A new template TF relation has been derived, as presented in Paper I.

We have also used the 2MRS density field to correct our measured distances and peculiar velocities for Malmquist bias.  The corrections require us to construct ad hoc selection criteria for our sample, and we strongly advise that if one is to make use of a subset of our data that involves imposing additional selection constraints, then one should not use the same bias corrections presented here.  Additionally, our Malmquist bias corrections are likely to be of limited value for galaxies with distances close to or greater than 20,000 \kms.  (Though very few galaxies in our sample are found in this regime anyway.)

The complete TF catalog is by far the largest such peculiar velocity catalog with $\sim 15\%$ distance errors to date.  SFI++ is also unique among peculiar velocity catalogs in that it includes a mix of cluster and field galaxies, as well as group identifications for galaxies in loose groups.

The detailed spectroscopic datasets, including parameters not directly used in the computation of TF distances, have been presented in S05 and Catinella, Haynes, \& Giovanelli (2005), but we have provided the detailed photometric parameters for most of the galaxies in our sample here, including galaxies for which we do not have high quality width measurements, which are therefore excluded from the TF sample.  The distribution of parameters is very similar to those provided in subsamples of the data published by our group before we made recent changes to the data processing algorithms.  The entirety of the TF dataset is available online through NVO-compliant tables hosted by the Cornell Theory Center.

\vskip 0.3in

We wish to thank our numerous collaborators who have participated over many years.  We especially thank Pirin Erdogdu for providing a copy of the 2MASS density field and Barbara Catinella for her work in deriving the relations between optical and radio line widths.  This work has been partially supported by NSF grants AST-9900695, AST-0307396, and AST-0307661.  CMS was supported by the N.R.A.O./GBT 03B-007 Graduate Student Support Grant and the NASA New York State Space Grant while at Cornell, and now holds a National Research Council Research Associateship Award at the Naval Research Laboratory.  Basic research in astronomy at the Naval Research Laboratory is funded by the Office of Naval Research.  KLM is a Harvard Postdoctoral Research Fellow supported by NSF grant AST-0406906.

\vskip 0.3in

{\bf Tables 1, 2, 3, and 4} are exclusively available online at {\it http://arecibo.tc.cornell.edu/hiarchive/sfiplusplus.php}.

\vfill
\end{document}